# Two-dimensional topological materials discovery by symmetry-indicator method

Di Wang[†,1,2], Feng Tang[†,1,2], Jialin Ji[3], Wenqing Zhang[4], Ashvin Vishwanath[5], Hoi Chun Po[6], Xiangang Wan[*,1,2]

[1]National Laboratory of Solid State Microstructures and School of Physics, Nanjing University, Nanjing 210093, China

[2]Collaborative Innovation Center of Advanced Microstructures, Nanjing 210093, China

[3]Materials Genome Institute, Shanghai University, Shanghai 200444, China

[4]Department of Physics and Shenzhen Institute for Quantum Science & Technology, Southern University of Science and Technology, 1088 Xueyuan Road, Shenzhen, Guangdong 518055, China

[5]Department of Physics, Harvard University, Cambridge, MA, USA

[6]Department of Physics, Massachusetts Institute of Technology, Cambridge, MA, USA

Two-dimensional (2D) topological materials (TMs) have attracted tremendous attention due to the promise of revolutionary devices with non-dissipative electric or spin currents. Unfortunately, the scarcity of 2D TMs holds back the experimental realization of such devices. In this work, based on our recently developed, highly efficient TM discovery algorithm using symmetry indicators, we explore the possible 2D TMs in all non-magnetic compounds in four recently proposed materials databases for possible 2D materials. We identify hundreds of 2D TM candidates, including 205 topological (crystalline) insulators and 299 topological semimetals. In particular, we highlight MoS, with a mirror Chern number of -4, as a possible experimental platform for studying the interaction-induced modification to the topological classification of materials. Our results winnow out the topologically interesting 2D materials from these databases and provide a TM gene pool which for further experimental studies.

## Introduction

The past decade has seen a rapid growth of research interest in topological materials (TMs) in which the exotic electronic states may provide a new avenue for spintronics, quantum computation and many other potential device applications [1,2]. Currently, a large number of weakly correlated three-dimensional (3D) TMs have been theoretically proposed and some of them have been experimentally confirmed, including topological insulators (TIs) [1,2], topological crystalline insulators (TCIs) [3], topological semimetals (TSMs) [4,5], and various other topological phases with more refined differences in fermiology [6-13].

Compared to their 3D counterparts, two-dimensional (2D) TMs could be better suited for technological applications due to their reduced dimensionality, especially in devices utilizing coherent spin transport [1,2,14]. For instance, the one-dimensional topologically protected edge states of 2D TMs should be more immune to undesired backscattering. Furthermore, 2D materials can be readily assembled into a myriad of heterostructures with highly tunable physical properties [15,16]. However, very few 2D TIs have been experimental realized yet despite the large number of theoretical proposals [14,17-22]. The well-known examples are HgTe/CdTe and InAs/GaSb/AlSb quantum well systems [23,24]. Unfortunately, due to a small nontrivial bulk band gap it is challenging to leverage their topological characters for technological applications. Recently, signatures of the quantum spin Hall effect in thin film $WTe_2$ have also been reported [25,26]. But $WTe_2$ has significant elastic scattering in the edge, which hampers its applications [25]. In addition, even less experimental progress has been made on 2D TCI [27] and TSM [28].

Such scarcity for experimental materials platforms can be partly attributed to the fact that the study of 2D materials is a relatively young field, with most of the recent development stimulated by the mechanical exfoliation of graphene in 2004 [29]. Since then, research interest in studying 2D materials has grown exponentially and dozens of 2D materials have been successfully synthesized [15,16], with notable examples being the graphene family with honeycomb structures, transition metal dichalcogenides, and metal halides. To further expand on the families of 2D materials, several databases for potential 2D crystals were recently developed by either studying the prospect of exfoliating atomically thin layers from 3D parent compounds [30-32], or by the combinatorial lattice decoration of known crystal structure prototypes [33]. Compared with the early 2D materials databases which are scarce and less developed, these four databases are more comprehensive and provide a basis for the large-scale prediction for 2D TMs.

In this work, we perform a comprehensive search for 2D TMs in these databases. While TMs discovery using conventional wave-function-based method is computationally costly, we circumvent this difficulty by adopting a symmetry-based perspective---an approach that has recently been successfully applied to 3D TMs discovery [34-36]. More specifically, we apply the highly efficient algorithm [37] based on the theory of symmetry indicators (SIs) of band topology [38], which allows for the detection of TIs, TCIs, and TSMs in a single calculation, to perform a systematic search for TMs candidates. Of the 3471 entries in these four databases [30-33], we found that 7.7% of them are candidates for T(C)Is, and 11.2% for TSMs[1]. Note that, there are some unavoidable overlaps across the entries in these different 2D databases, i.e., some nominally different entries only differ in a small difference in lattice constants or atomic positions. Such overlapping entries are particularly promising for they are identified through independent methods. We provide a list of all the 2D TM candidates in the SM, in which materials entries with almost the same structure are considered as a single candidate in counting. In particular, we identify 20 TIs with a full band gap of at least 25 meV, the scale of room temperature. In addition, we identify MoS as a potential material realization of a TCI with a mirror Chern number of -4, which could provide a first experimental platform for the study of how strong electron-electron interactions can modify the topological classification of materials [39]. Our results provide a topological characterization of the potential 2D materials in the mentioned databases [30-33], which could guide the experimental realization of new families of 2D topological materials.

[1] Note that of the 3471 2D entries, we find 2664 different materials after accounting for duplicates. Among them, 1835 materials are found to be non-magnetic. Thus, the proportions for T(C)Is and TSMs with respect to the non-magnetic materials are 11.2% and 16.3%, respectively.

## 2D topological phases and symmetry indicators

In the presence of spin-orbit coupling (SOC), a 2D nonmagnetic topological (crystalline) insulator with stable 1D edge modes is essentially characterized[2] by a nontrivial $Z_2$ Kane-Mele index and/or a mirror Chern number (MCN) [1-5]. The MCN is well-defined as long as there is a mirror symmetry with respect to the plane parallel to the crystal. For non-interacting electrons, the MCN can take any integer value like the Chern number; however, in the presence of electron-electron interactions, insulators with MCN = ±4, ±8, … become smoothly connected to the trivial phase [39]. In this work, we focus exclusively on 2D crystalline materials whose symmetries can be described by one of the 80 layer groups (LGs), and apply the theory of symmetry indicators of band topology [37,38] to efficiently identify TM candidates from the symmetry properties of their Bloch wave functions. For brevity, in the following we only provide a quick review of the key ideas behind the diagnosis, and relegate the more elaborated discussions to the Methods and SM.

Given any crystalline material, one can compute how the electronic Bloch wave functions transform under the spatial symmetries at the high-symmetry momenta. As demonstrated in Refs. [38,40], this data can be succinctly summarized into an integer-valued vector, which can then be expanded with respect to the bases computed from the set of all possible atomic insulators with the same spatial symmetries [37,38]. In other words, to any crystalline material we can associate a set of expansion coefficients **q** [37]. The symmetry indicator (SI) is readily obtained from **q** (Methods) and allows one to identify materials that are definitely nontrivial. Given a fixed symmetry group, the SIs form an abelian group which we denote by $X_{BS}$, and a non-zero SI indicates a nontrivial topological (crystalline) insulator. In Table I, we summarize the relation between $X_{BS}$ and the 2D topological indices. In particular, if the MCN of the crystal can be defined, the $Z_2$ TI index is equivalent to the parity of the MCN [27]. Correspondingly, if $X_{BS}$ is $Z_2$, $Z_4$, or $Z_6$, the odd SI corresponding to a TI phase with an odd MCN, while the nonzero even SI indicate a TCI phase with an even MCN but not a $Z_2$ TI index (Table I); however, if $X_{BS}$ is $Z_3$, the parity of the MCN cannot be diagnosed from the SI alone. For such cases, one needs a further calculation of the MCN to extract the $Z_2$ index.

In the following, we will discuss several typical representative TMs candidates in details to illustrate how 2D TMs can be identified using the symmetry indicators.

[2] Recently proposed 2D high-order TCI [11-13] may also host in-gap corner states. However, without particle-hole symmetry the energies of these corner states are not protected, and so they may not be distinguishable from the bulk states. These systems are not considered as TCIs in our present context [38].

| $X_{BS}$ | SI | ν | MCN | typical material |
|---|---|---|---|---|
| $Z_2$ | 1 | 1 | odd (if defined) | $WTe_2$(LG15) [19], WO(LG64), ZrO(LG64) |
| $Z_3$ | 1,2 | 1 | odd | WO(LG78), $PdSe_2$(LG78), $PdS_2$(LG78) |
| | | 0 | nonzero and even | MoS(LG78) |
| $Z_4$ | 1,3 | 1 | odd | CdTe(LG61), ZnTe(LG61) |
| | 2 | 0 | nonzero and even | SnTe(LG61) [41] |
| $Z_6$ | 1,3,5 | 1 | odd | graphene(LG80)[29], $Si_3P$(LG80) |
| | 2,4 | 0 | nonzero and even | MoN(LG80) |

Table I. The detailed topological classification for 2D topologically nontrivial insulators, where ν is $Z_2$ topological invariant and MCN denotes mirror Chern number for the 2D material owning a mirror symmetry plane which coincides with the 2D layer. 2D TI corresponds to cases with $\nu = 1$ while 2D TCI corresponds to cases with MCN being any nonzero even number. We list several typical materials with relatively clean Fermi surfaces in the table. Materials analyzed in the main text are indicated in blue; several well-known examples are also listed in the table.

## 2D topological insulators

We first discuss three materials crystallizing in LG 78: WO, $PdSe_2$ and $PdS_2$. They are non-centrosymmetric and so their topological characters cannot be diagnosed through the Fu-Kane parity criterion [42]. Nevertheless, the SI method allows us to efficiently uncover their nontrivial band topology.

Fig. 1 Electronic band plots for the typical topological insulators: (a,b) WO , (c,d) $PdS_2$ and (e,f) $PdSe_2$ crystallizing in LG 78. The left panel corresponds to the GGA calculations while the right panel corresponds to the HSE06 calculations.

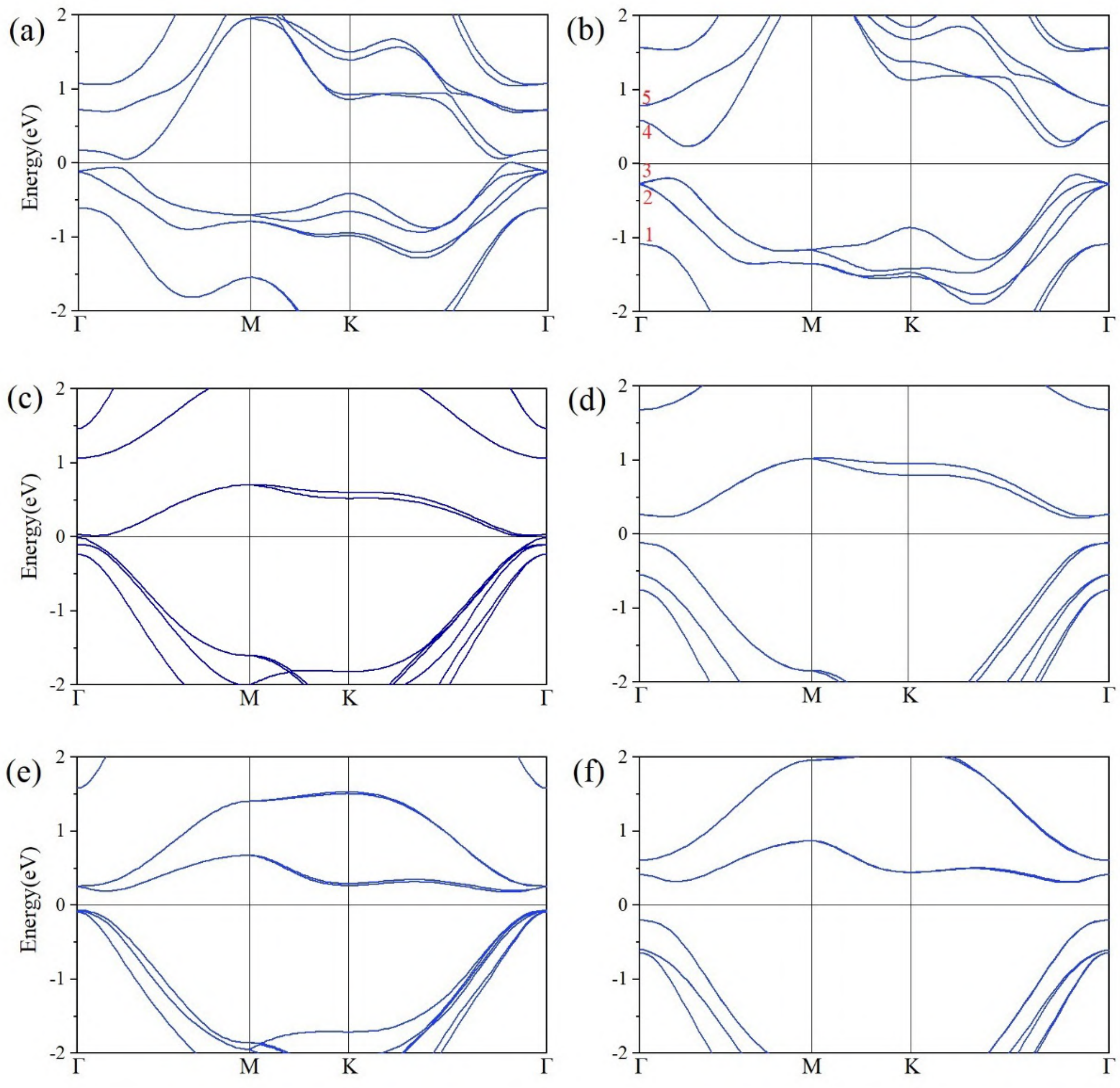

WO crystallizes in LG 78 with a hexagonal structure, and has two formula units in the primitive unit cell. There are three high symmetry points (HSPs) in the Brillouin zone for LG 78, which are Γ, M and K. The dispersion along all the high-symmetry paths from GGA calculation is shown in Fig. 1(a). Our numerical results show that O 2p orbitals are mainly located at the energy range -8 to -4 eV, the energy range near Fermi level is dominated by W 5d states, and this material has a full band gap of around 44 meV. As shown in Fig. 1(a), the direct gaps at M and K points in WO are relatively big and there is a likely band inversion between the upper and lower states around the Fermi level at Γ point. Usually, standard GGA calculations underestimate the band gap, thus we also perform HSE06 calculations to cross-check the results. Though band structures from GGA and HSE06 calculations for WO have some differences, most notably in the sizes of band gaps, we find the same diagnosis on band topology from the two methods. As shown in Fig. 1(b), the results from HSE06 calculation strengthen the band dispersion but do not change the band ordering. The estimated value of band gap from HSE06 calculation is 375 meV, while the strength of the band inversion at Γ point is also enhanced.

To prove that WO is topologically nontrivial beyond the heuristics of band inversion, we characterize it using the method of symmetry indicators [37]. LG78 has 7 AI basis vectors (i.e.

$d_{AI}$ = 7) with the last one owning a common factor 3, and so the SI group $X_{BS}$ is $Z_3$ [38]. Based on the wave functions from GGA and HSE06 calculations, we calculate the numbers of occurrences of all the irreducible representations (irreps) of little group for all the occupied bands at the three HSPs, and obtain the vector **n** (See Methods). Then we expand **n** with respect to these 7 AI basis vectors. The obtained expansion coefficients from GGA and HSE06 calculations are both **q** = (4, 0, 0, 1, 2, 1, 2/3). The last coefficient of 2/3 translates into SI = 2 in $Z_3$, which implies WO must have a nonzero MCN [43,44].

We also extend the SI analysis to study the effect of changes in the band ordering near the Fermi level, which allows us to assess the robustness of our prediction. As shown in Fig. 1(a), the bands at M and K points are far away from the Fermi level and therefore we focus on the bands near Fermi level at Γ point. Due to time-reversal symmetry, all the bands are doubly degenerate at Γ point. As shown in Fig. 1(b), we label the bands near the Fermi level at Γ point by 1, 2, 3, 4 and 5 in order of their energies, where bands 2 and 3 have almost the same energy. The point group of Γ in LG 78 is $D_{3h}$ and the bands labelled by 1, 2, 3, 4 and 5 are found to correspond to three different irreps, where the 2nd and 5th bands share the same irrep, and the 3rd and 4th bands share another irrep (see Table I in SM). Note that while an external field or strain may change the band ordering near the Fermi level, a band switching between the top of valence band and the bottom of conduction band (i.e. the 3rd and 4th bands in Fig. 1(b)) does not change the SI of the material, implying it must remain topological. Furthermore, a band switching between the 2nd the 4th bands would turn this material to be another topological phase with SI = 1 (see Table III in SM). Considering all the possible switches in the labelled bands, we find that WO could have a trivial SI only if the band switching occurs between the 1st and 4th bands, or the 3rd and 5th bands, both of which are unlikely under small perturbations. Thus, the topological nature of this material is likely robust to the presence of external field or strain.

Within the same LG 78, we find that $PdX_2$ (X = S, Se) also have SI = 2 in $Z_3$. This indicates that they are topological (crystalline) insulators with nonzero MCNs. They again take a hexagonal structure with two formula units in each primitive unit cell. We perform the GGA calculation for $PdS_2$, and show its band structures in Fig. 1(c). The bands around Fermi energy mainly come from Pd 4d states and the band gap is estimated to be about 13 meV. As shown in Fig. 1(d), HSE06 calculations strengthen the band dispersions and enlarge the splitting in Gamma point around Fermi level for $PdS_2$. The results from HSE06 calculations predict a larger gap of about 337 meV with the topological properties unaltered. For $PdSe_2$, which can be considered as substituting S ions in $PdS_2$ by Se ions, the band gap could be even larger. As shown in Fig. 1(e) and 1(f), the band gap from GGA calculation for $PdSe_2$ is 246 meV while HSE06 calculation predicts a much larger value of 506 meV. Though $PdSe_2$ has a much larger band gap than $PdS_2$, the band dispersions for these two materials are similar.

While the SI method allows us to efficiently prove that the MCNs WO and $PdX_2$ are nonzero, as shown in Table 1 and discussed in the preceding section, we cannot determine if they are TIs protected solely by time-reversal symmetry, or are TCIs protected by the mirror. To this end, we compute their MCNs using the hybrid Wannier center method [45], and found them to be -1 for all three materials (see SM for details). This implies they are all 2D TIs.

We also consider LGs with $X_{BS}$ being $Z_2$, $Z_4$ or $Z_6$. For such LGs, one can directly distinguish TIs from TCIs simply by the parity of the SI as shown in Table I. We reiterate that the parity of the SI in these cases is essentially identical to the Fu-Kane criterion [42]. Based on GGA calculations, we find that XO (X=W, Zr) crystallizing in LG 64 are 2D TIs, whose SI is found to be 1 within the $Z_2$ SI group. In the case of $Z_4$ SI group, we find some 2D TIs such as CdTe and ZnTe crystallizing in LG 61. In the case of $Z_6$ SI group, $Si_3P$ crystallizing in LG 80 is found to be a 2D TI with SI being 1. We also perform HSE06 calculations for these materials, and the results of topological properties are consistent with the standard GGA calculations. As a sanity check, we note that the well-known 2D TIs graphene (with SOC) [29] and $WTe_2$ series [19] are also identified as TIs in our algorithm (Table I).

## 2D topological crystalline insulators

Apart from TIs, 2D TCIs can also be discovered using the SI method [37]. In our context, such TCIs are essentially materials with even MCNs. We reiterate that, here, we focus on TCIs with robust symmetry-protected edge states. As shown in Table I, these TCIs can be uncovered when SI group of the corresponding LG is $Z_3$, $Z_4$ or $Z_6$, but for the $Z_3$ case one has to further evaluate the MCN to ensure it is even.

Fig. 2 Electronic band structures for typical 2D topological crystalline insulator MoS in LG 78: (a) GGA calculated band structure (b) HSE06 calculated band structure.

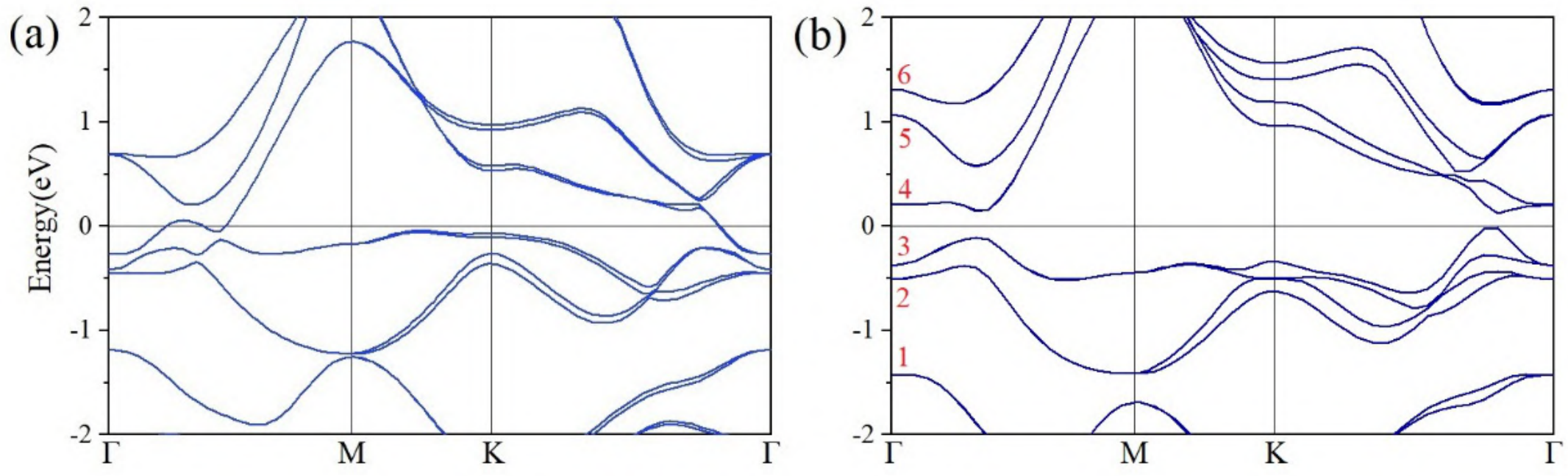

We first discuss an example material candidate in LG 78, whose SI group is $Z_3$. MoS crystallizes in a hexagonal structure and have two formula units in each primitive unit cell. We perform GGA calculation and show the electronic band structures in Fig. 2(a). GGA calculation suggests that MoS has a continuous direct gap everywhere in the Brillouin zone. The bands around the Fermi energy mainly come from Mo-4d states, whereas the S-3p states are mainly located between -6 and -2 eV. However, due to strong hybridization with the Mo-4d states, the S-3p states also show significant contribution to states at 3 eV above the Fermi level. To crosscheck the results, we also perform HSE06 calculations. As shown in Fig 2(b), within the HSE06 scheme this material is predicted to be an insulator with a full gap about 140 meV. The HSE06 calculation suggests stronger dispersion but does not change the band ordering. Both our GGA and HSE06 calculations give the expansion coefficients **q** = (4, 0, 0, 1, 2, 1, 2/3) (See SM for all the AI basis vectors), which again translates into a nonzero SI of 2 in $Z_3$. Similar to the discussion above for WO in LG 78, we study the topological features by analyzing the bands at Γ point. From an analysis on how a change in the band ordering might affect the SI (see SM for details), one concludes that MoS is also a robust TM with respect to small perturbations. To further verify that MoS is a TCI instead of a TI, we compute its MCN within GGA calculation, and found it to be -4 (see SM for details). This highlights MoS as a particularly interesting material for the study on the interplay between electron-electron correlations and the topological classification of phases of matter [39].

We also discovered several TCIs within LGs with the SI group being $Z_4$ or $Z_6$. Here, we highlight two particular candidates. First, we find a SI of 2 in $Z_4$ for the SnTe series crystalizing in LG 61. This is consistent with the previous characterization of the system as a 2D TCI with even MCN [41]. Second, we also identify MoN in LG 80 as a TCI from its SI of 4 in $Z_6$.

## 2D topological semimetals

Fig. 3 Electronic band plots for the 2D topological Dirac semimetal $TaBr_2$ crystallizing in LG 15. GGA calculation is shown in (a) and the HSE06 calculation is shown in (b).

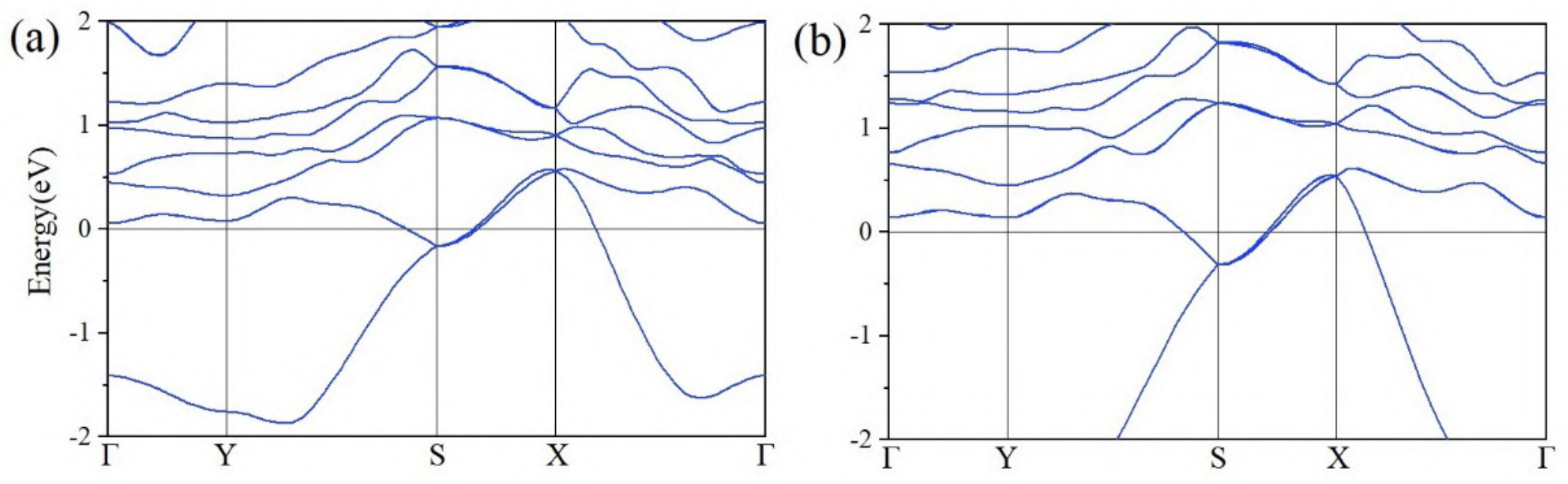

The SI method is equally well-suited for the discovery of TSMs [37]. Generally, the (4-fold degenerate) Dirac points require protection from crystalline symmetries. Unlike in 3D systems, 2D materials naturally have a smaller set of crystalline symmetries, and thus 2D Dirac materials candidates are rare when SOC is considered. Here, we identify $TaX_2$ (X = Br, I) as a 2D Dirac semimetal with non-negligible SOC. $TaX_2$ (X = Br, I) crystallizes in LG 15 with a rectangular structure, and has two formula units in the primitive unit cell. The electronic band structure for $TaBr_2$ is shown in Fig. 3. The bands around the Fermi level are mainly coming from Ta 5d orbitals. Four-fold band crossings, protected by a non-symmorphic 2-fold screw rotation ($\widetilde{C_{2x}}$) and inversion symmetries**,** are found near the Fermi surface at momenta S and X[3]. We remark that, in fact, $TaX_2$ is a filling-enforced semimetal [46] as the states at X and S are all four-fold degenerate, and there are 38 valence electrons per unit cell. Further symmetry analyses show that energy

[3] Let us illustrate it for the S point, and the argument for the X point is similar: let $\psi$ be an energy eigenstate at S with a definite parity, then one can show that the degenerate states $T\psi, \widetilde{C_{2x}}\psi, T\widetilde{C_{2x}}\psi$ ($T$ is time-reversal operator) are all orthogonal to $\psi$ and one another.

dispersion around S or X is linear in all directions.

## Conclusions and Discussions

To conclude, we have found 205 T(C)Is and 299 TSMs, among which 20 TIs have full band gaps that are larger than 25 meV, the energy scale corresponding to room temperature. These candidates are listed in the SM alongside with the computed band gaps. Note that, as shown in the SM, some overlapping entries across the databases [30-33] are consistently identified to be topological in spite of their slight structural differences. These materials are particularly promising candidates and deserve future comprehensive study. We also list 12 small gap (< 50 meV) ordinary insulator materials in the SM since 2D system is sensitive to external perturbation so that their physical properties and topological characters could be changed by the application of a small external field or strain, a desired feature for functional materials. In addition, we identify MoS as a possible candidate for the experimental study of how electron-electron interactions can affect the topological classification of phases of matter. To aid the SI analysis of any future 2D materials, we also provide the explicit AI basis vectors for the 80 LGs in the SM. We expect the 2D materials candidates highlighted in this work to provide a gene pool which could expedite the integration of 2D TMs in to functional devices.

## Acknowledgements

DW, FT and XW were supported by the NSFC (No.11525417, 11834006, 51721001 and 11790311), National Key R&D Program of China (No. 2018YFA0305704 and 2017YFA0303203) and the excellent programme in Nanjing University. DW was also supported by Program B for outstanding PhD candidates of Nanjing University. AV was supported by a Simons Investigator Award. HCP was supported by a Pappalardo Fellowship at MIT and a Croucher Foundation Fellowship. W.Z. acknowledges the support from Guangdong Innovation Research Team Project (Grant No. 2017ZT07C062), and Shenzhen Pengcheng-Scholarship program.

## Methods

The first-principles calculations are performed by using the projector-augmented wave method and a plane wave basis set as implemented in the Vienna ab initio simulation package (vasp) [47].

The Perdew–Burke–Ernzerhof (PBE) of generalized gradient approximation (GGA) is chosen as the exchange-correlation functional [48]. In our calculations, the cutoff energy of planewaves is set as 500 eV, and a k-mesh of 2D Brillouin zone is set by $\frac{60}{a}$ along each direction, where $a$ denotes the length of the lattice constant in Å. The structure file for the materials are made manually by separating single layers and adding a 30 Å vacuum padding in the z-direction. We crosscheck the results in the main text obtained from the standard GGA calculation against those from Hybrid functional calculation (HSE06) [49]. The effect of spin-orbit coupling (SOC) is considered self-consistently in all the calculations.

We first screen out the 1189 magnetic materials entries from the databases [30-33] with convergent total magnetic moment larger than 0.01 $\mu_B$, and then apply the SI algorithm [37,38] to the remaining 2282 non-magnetic 2D entries and extract topological ones in a single sweep by the following steps:

1. First, we generate the atomic insulator (AI) basis vectors: there are 80 layer groups (LGs) in total. For each LG, we construct the corresponding AI basis vectors: $\boldsymbol{a}_i, i = 1,2,\dots,d_{\mathrm{AI}}$, where $d_{\mathrm{AI}}$ is the number of AI basis vectors [38]. The common factor for each AI basis vector $\boldsymbol{a}_i$ is denoted by $\mathrm{C}_i$. We arrange the AI basis vectors in ascending order of their common factors. For LGs, only the last common factor is larger than 1, and the SI group can be written as $\mathrm{X_{BS}} = Z_{C_{d_{\mathrm{AI}}}}$ [38].
2. For each 2D material, by the first-principles calculated results and its LG, we calculate $\mathbf{n} = \left(\nu_{\mathrm{e}}, \mathrm{n}_{\mathrm{k}_1}^1, \mathrm{n}_{\mathrm{k}_1}^2, \dots, \mathrm{n}_{\mathrm{k}_2}^1, \mathrm{n}_{\mathrm{k}_2}^2, \dots\right)$ from the first $\nu_{\mathrm{e}}$ bands. Here, $\nu_{\mathrm{e}}$ is the total number of valence electrons per primitive unit cell. The other integers $\mathrm{n}_{\mathrm{k_i}}^{j}$ are the counts of the j-th irreducible representation of the little group $\mathrm{G(k_i)}$ at the i-th high-symmetry point $\mathrm{k_i}$.
3. Finally, we expand $\mathbf{n}$ on the AI basis of this LG by $\mathbf{n} = \sum_{i=1}^{d_{\mathrm{AI}}} q_i \boldsymbol{a}_i$. If the $q_i$ are all integers, the system could be topologically trivial. *If there exists some $q_i$ which is not an integer, then this material must be topological*, and furthermore, if all the $(\mathrm{q_i}C_i)$'s are integers, the material is a TI/TCI owning integer SI being $\mathrm{q_i C_i} \bmod \mathrm{C_i}$ for $\mathrm{C_i} > 1$ (the corresponding SI of any $\mathrm{C_i}$=1 must be 0), otherwise it is a TSM [37].

# Supplementary Materials for "Two-dimensional topological materials discovery by symmetry-indicator method"

Di Wang,[1,2,*] Feng Tang,[1,2,*] Jialin Ji,[3] Wenqing Zhang,[4]
Ashvin Vishwanath,[5] Hoi Chun Po,[6] and Xiangang Wan[1,2,†]

[1] *National Laboratory of Solid State Microstructures and School of Physics, Nanjing University, Nanjing 210093, China*
[2] *Collaborative Innovation Center of Advanced Microstructures, Nanjing University, Nanjing 210093, China*
[3] *Materials Genome Institute, Shanghai University, Shanghai 200444, China*
[4] *Department of Physics and Shenzhen Institute for Quantum Science & Technology, Southern University of Science and Technology, 1088 Xueyuan Road, Shenzhen, Guangdong 518055, China*
[5] *Department of Physics, Harvard University, Cambridge, MA, USA*
[6] *Department of Physics, Massachusetts Institute of Technology, Cambridge, MA, USA*

## CONTENTS

* These authors contributed equally to this work.
† The corresponding author: xgwan@nju.edu.cn.

## I. REVIEW OF SYMMETRY-INDICATOR THEORY FOR THE 2D SYSTEMS

Any group of electronic bands with finite direct gaps throughout the whole Brillouin zone (BZ) above and below, can be characterized by the following "band structure" (BS) vectors [1]:

$$\mathbf{n} = (\nu, n^1_{\mathbf{k}_1}, n^2_{\mathbf{k}_1}, n^3_{\mathbf{k}_1}, \ldots, n^1_{\mathbf{k}_2}, n^2_{\mathbf{k}_2}, \ldots), \tag{1}$$

where $\nu$ is the total number in this group of electronic bands, and for the subsequent numbers in the form of $n^j_{\mathbf{k}_i}$, the subscript $\mathbf{k}_i$ denotes the $i$th high symmetry point (HSP) and the superscript $j$ denotes the $j$th irreducible representation (irrep) of the little group of $\mathbf{k}_i$, $\mathcal{G}(\mathbf{k}_i)$. A subspace for the entire BS space is the atomic insulators (AIs), which has a real-space description and own the same dimension, $d_{\text{AI}}$, as the BS space [1]. The common factors, $C_i, i = 1, 2, \ldots, d_{\text{AI}}$, of the AI basis vectors, $\mathbf{a}_i, i = 1, 2, \ldots, d_{\text{AI}}$, contain all the information of the SI group $X_{\text{BS}} = \mathbb{Z}_{C_1} \times \mathbb{Z}_{C_2} \ldots \times \mathbb{Z}_{C_{d_{\text{AI}}}}$ [1]. For example, for LG 2, there are 5 AI basis vectors, whose common factors are 1,1,1,1,2, respectively (we arrange the AI basis vectors in the ascending order of the common factors). So $X_{\text{BS}}$ for LG 2 is equal to $\mathbb{Z}_1^4 \times \mathbb{Z}_2$, isomorphic to $\mathbb{Z}_2$. Different from the 3D situation, the nontrivial symmetry-indicator (SI) groups, $X'_{\text{BS}}s$, for all possible symmetries of the 2D systems, i.e. 80 layer groups (LGs), can only be one of the following four possibilities: $\mathbb{Z}_2$, $\mathbb{Z}_3$,$\mathbb{Z}_4$,$\mathbb{Z}_6$ [1]. In the first principles calculations, we first calculate $\mathbf{n}$ of the first $\nu_e$ bands ($\nu_e$ is the total number of valence electrons per primitive unit cell, i.e. the filling number). Then we can expand it on the AI basis vectors as the following:

$$\mathbf{n} = \sum_i^{d_{\text{AI}}} q_i \mathbf{a}_i. \tag{2}$$

When all the $q_i's$ are integers, the material may be an atomic insulator [2]. When some of $(C_i q_i)'s$ is found to be non-integers, the material is definitely a topological semimetal (TSM) since the compatibility relations are not satisfied [2]. For the rest cases, which must be topological (crystalline) insulator [2] and belong to the LGs with nontrivial $X_{\text{BS}}$, one can compute $r_i = (C_i q_i) \mod C_i$, which can only take, $0, 1, \ldots, C_i - 1$. For these cases, $r_{d_{\text{AI}}}$ must be one of $1, \ldots, C_i - 1$. For $C_{d_{\text{AI}}} = 2, 4, 6$, i.e. $X_{\text{BS}} = \mathbb{Z}_2, \mathbb{Z}_4, \mathbb{Z}_6$, when $r_{d_{\text{AI}}}$ is an odd number, the material is a topological insulator (TI), while when $r_{d_{\text{AI}}}$ is an even number, the material is a topological crystalline insulator (TCI) [3, 4]. Attention should be paid to $C_{d_{\text{AI}}} = 3$, i.e. $X_{\text{BS}} = \mathbb{Z}_3$, $r_{d_{\text{AI}}}$ is 1, 2, indicating non-vanishing mirror Chern number, but the parity of the MCN cannot be diagnosed from the SI alone. For such cases, one needs a further calculation of the MCN. When the MCN is odd, the material is a TI otherwise it is a TCI [3, 4].

## II. IRREDUCIBLE REPRESENTATIONS OF THE LITTLE GROUP OF Γ POINT FOR LG 78

With respect to irreps of the little group of Γ point for LG 78, we can just consider the point group, namely $D_{3h}$ which has three 2D irreps [5]. To give a more detailed account, we can label these irreps according to the eigenvalues of $S_3 = C_3\sigma_z$ which belongs to $D_{3h}$. Because $S_3^6 = \bar{E}$, thus the eigenvalue of $S_3$, $\lambda$, must satisfy $\lambda^6 = -1$ thus $\lambda$ can take six different values as $e^{\pm i \frac{(2l+1)\pi}{6}}$ where $l = 0, 1, 2$. Furthermore, the two eigensates of $S_3$ for each $l(= 0, 1, 2)$ are related by time-reversal symmetry. To be specific, $l = 0, 1, 2$ doublets correspond to the 1st,2nd and 3rd irrep, respectively.

| | | | | | |
|---|---|---|---|---|---|
| Band label | 1 | 2 | 3 | 4 | 5 |
| Which irrep | 2 | 1 | 3 | 3 | 1 |

TABLE I. For WO (LG 78), the 5 bands at Γ point which are labelled by $1, 2, \ldots, 5$ in the main text, are attributed which irreps they belong. For example, the 5th band belongs to the 1st irrep.

| | | | | | | |
|---|---|---|---|---|---|---|
| Band label | 1 | 2 | 3 | 4 | 5 | 6 |
| Which irrep | 2 | 3 | 1 | 3 | 1 | 2 |

TABLE II. For MoS (LG 78), the 6 bands at Γ point which are labelled by $1, 2, \ldots, 6$ in the main text, are attributed which irreps they belong. For example, the 1st band belongs to the 2nd irrep.

The switch in valence and conduction bands between these three irreps affects the value of SI (assume $n^j_{\mathbf{k}}{}'s$ for the rest HSPs are unchanged), which is shown in Table III in the following. For example, WO (LG 78) owns SI=2 as discussed in the main text. If the 2nd band and the 4th band are switched, according to Table I, this corresponds to $(\delta n_\Gamma^1, \delta n_\Gamma^2, \delta n_\Gamma^3) = (-1, 0, +1)$, which will result in a new SI being 1 (See the number in row 4 and column 4 in Table III). While for MoS (LG 78) which also has SI=2 in the original state, in order to discuss the topological feature, we label the bands around the Fermi level located at  point as 1, 2, 3, 4, 5 and 6, as shown in Fig. 2(b) in the main text. These bands at  point correspond three different irreps, as shown in Table II. According to their irreps, similar to the discussion in WO system, the band switch between the 3rd and 4th bands around the Fermi level would turn this material to another topological phase with SI=1. For all the switches in the labelled bands, only three switches would turn MoS to be a trivial atomic insulator, which are the switches between 1st and 4th bands, between 2nd and 5th bands, between 3rd and 6th bands, respectively. Therefore, MoS is also a robust topological material with respect to small perturbations of external field or stress.

| The original SI \ $(\delta n_\Gamma^1, \delta n_\Gamma^2, \delta n_\Gamma^3)$ | (+1,-1,0) | (0,+1,-1) | (-1,0,+1) | (-1,+1,0) | (0,-1,+1) | (+1,0,-1) |
|---|---|---|---|---|---|---|
| 0 | 2 | 2 | 2 | 1 | 1 | 1 |
| 1 | 0 | 0 | 0 | 2 | 2 | 2 |
| 2 | 1 | 1 | 1 | 0 | 0 | 0 |

TABLE III. $(\delta n_\Gamma^1, \delta n_\Gamma^2, \delta n_\Gamma^3)$ mean the variation of $n_\Gamma^j, j = 1, 2, 3$ after one irrep at $\Gamma$ point is occupied one more time while the occupation number for another irrep decreased by one. In the row indicated by "the original SI": 0,1,2, the values in the rest columns represent the resulting SI after the variation.

## III. GOOD 2D TOPOLOGICAL INSULATOR CANDIDATES WITH FULL GAP

We list all the relatively good topological insulator candidates with full gaps larger than room temperature from GGA calculations in the following table. Meanwhile, we give all the band gaps behind the names of these materials. We also list the corresponding database [6–9] behind the names of these materials. Note that there are some materials belong to different databases by a small structural difference, are found to belong to the same topological classification. Moreover, for the materials which has been proposed theoretically in previous work, we list the references behind their names.

TABLE IV. Good 2D topological insulator candidates with their band gaps

| $LG$ | $X_{\text{BS}}$ | Topological insulators |
|---|---|---|
| 15 | $\mathbb{Z}_2$ | $IrTaTe_4$ (0.033 eV) [7, 10], $MoS_2$ (0.048 eV) [9, 11], $MoSe_2$ (0.033 eV) [9, 11], $S_2W$ (0.056 eV) [9, 11], $Se_2W$ (0.033 eV) [9, 11] |
| 46 | $\mathbb{Z}_2$ | BrRhS (0.040 eV) [9], ClNPb (0.032 eV) [9], ClSeTl (0.027 eV) [9], $HfTe_5$ (0.115 eV) [6, 12], $HfTe_5$ (0.098 eV) [7, 12], INTi (0.044 eV) [6, 13], INTi (0.040 eV) [7, 13], INTi (0.030 eV) [8, 13], INTi (0.038 eV) [9, 13], $Te_5Zr$ (0.137 eV) [6, 12] |
| 47 | $\mathbb{Z}_2$ | $Ge_3N$ (0.077 eV) [9] |
| 72 | $\mathbb{Z}_2$ | Bi (0.538 eV) [7, 14], Bi (0.542 eV) [8, 14], FGe (0.098 eV) [9], FSn (0.284 eV) [9], Sn (0.073 eV) [9, 15] |
| 78 | $\mathbb{Z}_3$ | HgO (0.288 eV) [9], OW (0.044 eV) [9], $PdSe_2$ (0.246 eV) [9] |
| 79 | $\mathbb{Z}_3$ | $I_3Re$ (0.130 eV) [9], $I_3Re$ (0.139 eV) [9] |

## IV. THE PREDICTED 2D TOPOLOGICAL INSULATOR CANDIDATES

We list the topological insulator candidates from GGA calculations in the following table. We also list the corresponding database behind the names of these materials. Note that there are some materials belong to different databases by a small structural difference, are found to belong to the same topological classification. Moreover, for the materials which has been proposed theoretically in previous work, we list the references behind their names.

TABLE V. 2D topological insulator candidates

| $LG$ | $X_{BS}$ | Topological insulators |
|---|---|---|
| 14 | $\mathbb{Z}_2$ | $AuCuO_2$ [6], $AuKSe_2$ [7], $Se_4TiZr$ [6, 10], $Se_4TiZr$ [7, 10] |
| 15 | $\mathbb{Z}_2$ | $Cl_2Mo$ [9], $Cl_2W$ [9], $CrS_2$ [9, 11], $CrSe_2$ [9, 11], $GeTe_2$ [9], $HgTe_2$ [9], $IrTaTe_4$ [7, 10], $MoS_2$ [9, 11], $MoSe_2$ [9, 11], $MoSe_3$ [9, 11], $MoTe_2$ [6, 11], $MoTe_2$ [7, 11], $MoTe_2$ [8, 11], $MoTe_2$ [9, 11], $NiTe_3$ [9], $PbTe_2$ [9], $S_2W$ [9, 11], $Se_2W$ [9, 11], $Se_3Ta$ [6], $Te_2W$ [6, 11], $Te_2W$ [7, 11], $Te_2W$ [8, 11], $Te_2W$ [9, 11] |
| 17 | $\mathbb{Z}_2$ | $CoTaTe_2$ [6] |
| 18 | $\mathbb{Z}_2$ | ClZr [6], $PdSe_6Ta_2$ [6] |
| 41 | $\mathbb{Z}_2$ | $BiO_4Re$ [6], CaSi [7], CaSn [7] |
| 46 | $\mathbb{Z}_2$ | AgBrSe [9], AgClS [9], AgClSe [9], AgISe [9], BrCoO [9], BrCuSe [9], BrNPb [9], BrRhS [9], BrRhSe [9], BrSeTl [9], ClCoO [9], ClCoS [9], ClCuSe [9], ClNPb [9], ClNPd [9], ClRhS [9], ClRhSe [9], ClSeTl [9], CuISe [9], GeIN [9], $HfTe_5$ [6, 12], $HfTe_5$ [7, 12], IMoN [9], IMoN [9], INbSe [9], INTi [6, 13], INTi [7, 13], INTi [8, 13], INTi [9, 13], INW [9], IORh [9], IRhS [9], ISeTl [9], $Te_5Zr$ [6, 12] |
| 47 | $\mathbb{Z}_2$ | $AsC_3$ [9], $AsGe_3$ [9], $AsSn_3$ [9], $BiC_3$ [9], $BiGe_3$ [9], $BiSi_3$ [9], $Ge_3N$ [9], $Ge_3P$ [9], $Ge_3Sb$ [9], $NPb_3$ [9], $NSn_3$ [9], $PSn_3$ [9], $SbSi_3$ [9] |
| 61 | $\mathbb{Z}_4$ | CdSe [9], CdTe [9], SeZn [9], SZn [9], TeZn [9] |
| 64 | $\mathbb{Z}_2$ | GeTeZr [6, 16], GeTeZr [7, 16], HfS [9], HfSe [9], HfSiTe [6, 16], HfSiTe [7, 16], HfSiTe [8, 16], MoO [9], NiS [9], OOs [9], ORu [9], OsS [9], OsSe [9], OsTe [9], OTi [9], OW [9], OZr [9], PtS [9], PtSe [9], PtTe [9], SeTi [7], SeTi [9], SeZr [9], SiTeZr [8, 16], SZr [9], TeZr [9] |
| 72 | $\mathbb{Z}_2$ | Bi [7, 14], Bi [8, 14], BrZr [6, 17], BrZr [8, 17], $CF_2Mo_2$ [9], $CF_2Nb_2$ [9], $CF_2Ta_2$ [9], $CF_2V_2$ [9], $CF_2W_2$ [9], $CH_2Mo_2O_2$ [9], $CH_2Nb_2O_2$ [9], $CH_2O_2Ta_2$ [9], $CH_2O_2V_2$ [9], $CH_2O_2W_2$ [9], ClHLu [6], ClHLu [7], ClHLu [8], ClHSc [6], ClZr [7, 17], ClZr [8, 17], $CMo_2O_2$ [9], $CO_2W_2$ [9], $CS_2Ta_2$ [6], $CS_2Ta_2$ [7], $CS_2Ta_2$ [8], $CW_2$ [9], FGe [9], FSn [9], Ge [9, 18], GeILa [6], GeILa [8], $OsSe_2$ [9], $OsTe_2$ [9], SbTe [7], $Se_2Ti$ [6], $Se_2Ti$ [6], $Se_2Ti$ [7], $Se_2Ti$ [8], $Se_2Ti$ [9], Si [9, 18], Sn [9, 15], $Te_2Zr$ [8], $Te_2Zr$ [9] |
| 80 | $\mathbb{Z}_6$ | C [6, 19], C [8, 19], C [9, 19], $PSi_3$ [9] |

## V. THE PREDICTED 2D TOPOLOGICAL CRYSTALLINE INSULATOR CANDIDATES

We list the topological crystalline insulator candidates from GGA calculations in the following table. We also list the corresponding database behind the names of these materials. Note that there are some materials belong to different databases by a small structural difference, are found to belong to the same topological classification. Moreover, for the materials which has been proposed theoretically in previous work, we list the references behind their names.

TABLE VI. 2D topological crystalline insulator candidates

| $LG$ | $X_{\rm BS}$ | Topological crystalline insulators |
|---|---|---|
| 61 | $\mathbb{Z}_4$ | GeS [9, 20], GeSe [9, 20], GeTe [9, 20], GeTe [9, 20], PbS [9, 20], PbSe [9, 20], PbTe [9, 20], SeSi [9, 20], SeSn [9, 20], SiTe [9, 20], SnTe [9, 20], SSn [9, 20] |
| 80 | $\mathbb{Z}_6$ | MoN [7] |

## VI. THE PREDICTED 2D TOPOLOGICAL (CRYSTALLINE) INSULATOR CANDIDATES

As mentioned in the main text, for LGs with XBS being Z3, the nonvanishing SI, i.e. 1 and 2, must imply finite MCN, which may be an even or odd number, corresponding to a TCI or TI phase, with the Z2 topological invariant being 0 or 1, respectively; Hence, for this case one needs a further calculation of the Z2 topological invariant or MCN. We list all the topological (crystalline) insulator candidates for this case from GGA calculations in the following table. We also list the corresponding database behind the names of these materials. Note that there are some materials belong to different databases by a small structural difference, are found to belong to the same topological classification. Moreover, for the materials which has been proposed theoretically in previous work, we list the references behind their names.

TABLE VII. 2D topological (crystalline) insulator candidates

| $LG$ | $X_{\rm BS}$ | Topological crystalline insulators |
|---|---|---|
| 78 | $\mathbb{Z}_3$ | AuO [9], AuS [9], AuSe [9], AuTe [9], BiSn [6], $Br_2W$ [9], $C_2F_2Mo_3$ [9], $C_2F_2Ti_3$ [9], $C_2F_2W_3$ [9], $C_2Mo_3$ [9], $C_2Mo_3O_2$ [9], $C_2W_3$ [9], $Cl_2W$ [9], CuO [9], CuS [9], CuSe [9], CuTe [9], $GeTe_2$ [9], HfO [9], HgO [9], $HgS_2$ [9], $HgSe_2$ [9], $HgTe_2$ [9], MoO [9], MoS [9], MoSe [9], NiS [9], $NiS_2$ [9], NiSe [9], OOs [9], OPd [9], OPt [9], ORe [9], OsTe [9], OTa [9], OTi [9], OW [9], OZr [9], $PbTe_2$ [9], PdS [9], $PdS_2$ [9], PdSe [9], $PdSe_2$ [9], $PdTe_2$ [9], PtS [9], $PtS_2$ [9], PtSe [9], $PtSe_2$ [9], $PtTe_2$ [9], ReS [9], RuS [9], RuSe [9], RuTe [9], SeW [9], $SnTe_2$ [9], SW [9] |
| 79 | $\mathbb{Z}_3$ | $AgI_3$ [9], $Br_3Nb$ [9], $Br_3Ta$ [9], $Br_3Ta$ [9], $Cl_3Nb$ [9], $Cl_3Ta$ [9], $Cl_3Ta$ [9], $I_3Ir$ [9], $I_3Pt$ [9], $I_3Re$ [9], $I_3Re$ [9], $I_3Rh$ [9], $I_3Ta$ [9] |

## VII. ALL THE PREDICTED 2D TOPOLOGICAL SEMIMETAL CANDIDATES

We list all the topological 2D topological semimetal candidates from GGA calculations in the following table. We also list the corresponding database behind the names of these materials. Note that there are some materials belong to different databases by a small structural difference, are found to belong to the same topological classification. Moreover, for the materials which has been proposed theoretically in previous work, we list the references behind their names.

TABLE VIII. 2D topological semimetal candidates

| *LG* | topological semimetals |
|---|---|
| 9 | NS [8] |
| 15 | $AgBr_2$ [9], $AgI_2$ [9], $AlI_2$ [9], $AuBr_2$ [9], $AuI_2$ [9], $AuO_2$ [9], $AuTe_2$ [9], $Br_2Ga$ [9], $Br_2HZr_2$ [6], $Br_2In$ [9], $Br_2Ir$ [9], $Br_2Re$ [9], $Br_2Sc$ [9], $Br_2Ta$ [9], $Br_2Tl$ [9], $Br_2Tl$ [9], $Br_2Y$ [9], $Cl_2Ir$ [9], $Cl_2Nb$ [9], $Cl_2Re$ [9], $Cl_2Sc$ [9], $Cl_2Y$ [9], $CoTe_2$ [9], $CuI_2$ [9], $CuSe_2$ [9], $CuTe_2$ [9], $I_2In$ [9], $I_2Ir$ [9], $I_2Nb$ [9], $I_2Re$ [9], $I_2Sc$ [9], $I_2Ta$ [9], $I_2Y$ [9], $IrO_2$ [9], $IrS_2$ [9], $IrSe_2$ [9], $IrSe_3$ [9], $IrTe_2$ [9], $NbO_2$ [9], $NbSe_3$ [6], $NbSe_3$ [9], $NbTe_3$ [9], $Ni_2TaTe_3$ [6], $O_2Re$ [9], $O_2Rh$ [9], $O_2Ta$ [9], $ReS_2$ [9], $ReSe_2$ [9], $ReSe_3$ [9], $ReTe_2$ [9], $ReTe_3$ [9], $RhS_2$ [9], $RhSe_2$ [9], $RhTe_2$ [9], $S_2Sc$ [9], $S_3Sc$ [9], $S_3Ta$ [6], $S_3Ta$ [9], $ScSe_2$ [9], $ScSe_3$ [9], $ScTe_3$ [9], $Se_3Ta$ [9], $TaTe_3$ [9] |
| 17 | $AgF_2$ [8] |
| 28 | $NbPdTe_5$ [6] |
| 40 | $Br_3Ru$ [7] |
| 41 | Ga [7], $NbNiTe_5$ [6], $NiTaTe_5$ [6] |
| 46 | AgBrN [9], AgClN [9], AlBrN [9], AlClN [9], AlIN [9], AuBrN [9], AuClN [9], AuIN [9], BiBrN [9], BiClN [9], BiIN [9], BrCoN [9], BrGaN [9], BrGeO [9], BrGeS [9], BrGeSe [9], BrHfO [9], BrHfS [9], BrHfSe [9], BrHgS [9], BrHgSe [9], BrInN [9], BrIrN [9], BrMoO [9], BrMoS [9], BrMoSe [9], BrNiS [9], BrNiSe [9], BrNNb [9], BrNRh [9], BrNSc [9], BrNTa [9], BrOPb [9], BrOPd [9], BrOPt [9], BrORu [9], BrOSn [9], BrOsS [9], BrOsSe [9], BrOTi [8], BrOW [9], BrOZr [9], BrPbS [9], BrPbSe [9], BrPdS [9], BrPtS [9], BrPtSe [9], BrRuS [9], BrRuSe [9], BrSeSn [9], BrSeZr [9], BrSSn [9], BrSTi [9], BrSZr [9], ClCoN [9], ClCuN [9], ClCuN [9], ClGaN [9], ClGeO [9], ClGeS [9], ClGeSe [9], ClHfO [9], ClHfS [9], ClHfSe [9], ClHgS [9], ClHgSe [9], ClInN [9], ClIrN [9], ClMoO [9], ClMoSe [9], ClNiS [9], ClNiSe [9], ClNNb [9], ClNRh [9], ClNSc [9], ClNTa [9], ClOPb [9], ClOPd [9], ClOPt [9], ClORu [9], ClOSn [9], ClOsS [9], ClOsSe [9], ClOW [9], ClOZr [9], ClPbS [9], ClPbSe [9], ClPdS [9], ClPdSe [9], ClPtS [9], ClPtSe [9], ClRuS [9], ClRuSe [9], ClRuSe [9], ClSeSn [9], ClSeW [9], ClSeZr [9], ClSSn [9], ClSTi [9], ClSZr [9], CoIN [9], CuIN [9], CuTe [6], CuTe [8], GaIN [9], GeIO [9], GeIS [9], HfIO [9], HfIS [9], HfISe [9], HgIS [9], HgISe [9], IInN [9], IIrN [9], IMoO [9], IMoSe [9], INiS [9], INiSe [9], INNb [9], INRh [9], INSc [9], INTa [9], IOPb [9], IOPd [9], IOPt [9], IOSn [9], IOsS [9], IOsSe [9], IOTi [9], IOW [9], IOZr [9], IPbS [9], IPbSe [9], IPdS [9], IPdSe [9], IPtS [9], IPtSe [9], IRuS [9], IRuSe [9], ISeSn [9], ISeW [9], ISeZr [9], ISSn [9], ISTi [9], ISZr [9], $LaSeTe_2$ [6], $NbS_3$ [9] |
| 55 | $CuI_2O_2Sr_2$ [6] |
| 64 | CuSe [6], IrO [9], IrO [9], IrS [9], IrSe [9], IrTe [9], NbO [9], NbS [9], NbSe [9], NbTe [9], ORe [9], ORh [9], OSc [9], OTa [9], OV [9], ReS [9], ReSe [9], ReTe [9], RhS [9], RhSe [9], RhTe [9], ScSe [9], ScTe [9], SeTa [9], SSc [9], STa [9], TaTe [9], $Te_3Y$ [6], $Te_3Y$ [7], TeV [9], TeV [9] |
| 69 | SbTe [6], $Se_4Ta$ [7] |
| 72 | $CHf_2$ [9] |
| 78 | $AgI_2$ [9], $AlI_2$ [9], $AuI_2$ [9], $AuMg_3$ [7], $AuS_2$ [9], $AuSe_2$ [9], $AuTe_2$ [9], $Br_2Ga$ [9], $Br_2In$ [9], $Br_2In$ [9], $Br_2Nb$ [9], $Br_2Ta$ [9], $Br_2Tl$ [9], $C_2F_2Nb_3$ [9], $C_2F_2Ta_3$ [9], $C_2H_2O_2Ta_3$ [9], $C_2Nb_3$ [9], $C_2Nb_3O_2$ [9], $C_2O_2Sc_3$ [9], $C_2O_2Ta_3$ [9], $C_2O_2Y_3$ [9], $C_2Sc_3$ [9], $C_2Ta_3$ [9], $C_2V_3$ [9], $C_2Y_3$ [9], $Cl_2In$ [9], $Cl_2Nb$ [9], $Cl_2Ta$ [9], $Cl_2Tl$ [9], $Cl_2Tl$ [9], $CoSe_2$ [9], $CoTe_2$ [9], $CuI_2$ [9], $CuI_2$ [9], $CuSe_2$ [9], $CuTe_2$ [9], $GaI_2$ [9], $I_2In$ [9], $I_2Ir$ [9], $I_2Nb$ [9], $I_2Rh$ [9], $I_2Rh$ [9], $I_2Ta$ [9], $I_2Tl$ [9], $IrO_2$ [9], $IrS_2$ [9], $IrSe_2$ [9], $IrTe_2$ [9], $N_3W_2$ [8], $Ni_2SbTe_2$ [6], $Ni_2SbTe_2$ [7], $O_2Re$ [9], $ReS_2$ [9], ReSe [9], $ReSe_2$ [8], $ReSe_2$ [9], $ReTe_2$ [9], $RhS_2$ [9], $RhSe_2$ [9], $RhTe_2$ [9], $S_2Ta$ [6], $S_2Ta$ [7], $S_2Ta$ [8], $S_2Ta$ [9], $ScTe_2$ [9], $Se_2Ta$ [6] |

## VIII. THE SMALL GAP 2D TRIVIAL INSULATOR MATERIALS

TABLE IX. The small gap 2D trivial insulator materials

| *LG* | materials |
|---|---|
| 13 | PbSb (0.049 eV) [6] |
| 15 | $O_2Ru$ (0.030 eV) [9], $Se_3Ti$ (0.013 eV) [9] |
| 46 | BrIrSe (0.032 eV) [9], ClIrSe (0.031 eV) [9], ClORh (0.017 eV) [9] |
| 72 | $GeI_2Y_2$ (0.043 eV) [6] |
| 78 | $O_2Pb$ (0.022 eV) [9], TeW (0.021 eV) [9], $Te_2Ti$ (0.043 eV) [9], SeZr (0.033 eV) [9] |
| 79 | $AuCl_3$ (0.003 eV) [9] |

## IX. STATEMENT

Due to space limitations, the band structure plots of 2D topological materials, AI bases for 80 LGs and the details of mirror Chern number calculations are not shown in this version of Supplementary Materials. If you are interested in our work, please contact xgwan@nju.edu.cn